# DNA Translocation through Graphene Nanopores

Grégory F. Schneider, Stefan W. Kowalczyk, Victor E. Calado, Grégory Pandraud, Henny W. Zandbergen, Lieven M.K. Vandersypen and Cees Dekker*

*Kavli Institute of Nanoscience, Lorentzweg 1, 2628 CJ Delft, The Netherlands*

*Corresponding author: c.dekker@tudelft.nl

May 12, 2010

**Nanopores – nanosized holes that can transport ions and molecules – are very promising devices for genomic screening, in particular DNA sequencing[1,2]. Both solid-state and biological pores suffer from the drawback, however, that the channel constituting the pore is long, viz. 10-100 times the distance between two bases in a DNA molecule (0.5 nm for single-stranded DNA). Here, we demonstrate that it is possible to realize and use ultrathin nanopores fabricated in graphene monolayers for single-molecule DNA translocation. The pores are obtained by placing a graphene flake over a microsize hole in a silicon nitride membrane and drilling a nanosize hole in the graphene using an electron beam. As individual DNA molecules translocate through the pore, characteristic temporary conductance changes are observed in the ionic current through the nanopore, setting the stage for future genomic screening.**

In the past few years, nanopores have emerged as a new powerful tool to interrogate single molecules. They have been successfully used to rapidly characterize biopolymers like DNA[3,4], RNA[5], as well as DNA-ligand complexes[6] and local protein structures along DNA[7] at the single-molecule level. A key driving force for nanopore research in the past decade has been the prospect of DNA sequencing. However, a major roadblock for achieving high-resolution DNA sequencing with pores is the finite length of the channel constituting the pore (Fig. 1A). In a long nanopore, the current blockade resulting from DNA translocation is due to a large number of bases (for typical devices ~10-100 bases) present in the pore. Here, we demonstrate that this limitation can be overcome by realizing an ultimately thin nanopore in a graphene monolayer.

Graphene is a two-dimensional layer of carbon atoms packed into a honeycomb lattice with a thickness of only one atomic layer (~0.3 nm)[8]. Despite its minimal thickness, graphene is robust as a free standing membrane[9,10]. In addition, graphene is a very good electrical conductor[11]. Graphene therefore opens up new opportunities for nanopores such as new analytical platforms to detect, for example, local protein structures on biopolymers or sequencing with single-base resolution. Indeed, theoretical calculations of DNA translocation through a nanopore in graphene have already indicated the possibility for single-base resolution by probing the translocating molecule electrically in the transverse direction by use of the intrinsic conductive properties of graphene[12].

We obtain single-layer graphene (Fig. 1B) by mechanical exfoliation from graphite on $SiO_2$[13]. Monolayer graphene is identified by its particular optical contrast[14] in the optical microscope and by Raman measurements (Fig. 1C). At ~ 1590 $cm^{-1}$, we measure the so-called G resonance peak and at ~2690 $cm^{-1}$ the 2D resonance peak. In the case of multilayer graphene, the 2D resonance peak splits off in multiple peaks in contrast to monolayer graphene which has a very sharp single resonance peak. In this way, we are well able to distinguish single-layer graphene from multilayer graphene[15].

Next we select a monolayer of graphene and transfer it onto a SiN support membrane with a 5 micron sized hole[16] by use of our recently developed 'wedging transfer' technique[17]. This transfer procedure is straightforward: flakes can be overlaid to support membranes in less than an hour. Briefly, a hydrophobic polymer is spun onto a hydrophilic substrate (here plasma-oxidized $SiO_2$) with graphene flakes, and wedged off the substrate by sliding it at an angle in water. Graphene flakes are peeled off the $SiO_2$ along with the polymer. The polymer is then floating on the water surface, located near a target SiN substrate, the water level is lowered, and the flakes are positioned onto the SiN membrane with micrometer lateral precision. In the final step the polymer is dissolved.

We then drill a nanopore into the graphene monolayer using the highly focused electron beam of a transmission electron microscope (TEM). The acceleration voltage is 300 kV, well above the 80-140 kV knock-out voltage for carbon atoms in graphene[18] (see Methods). Drilling the holes by TEM is a robust well-reproducible procedure (we drilled 31 holes with diameters ranging from 2 to 40 nm, in monolayer as well as in multilayer graphene; some examples of pores are shown in Fig. 2). Because of the high acceleration voltage of the electron beam, drilling could potentially induce damage to the graphene around the pore. However, electron beam diffraction measurements across the hole (Fig. 2B and C) confirm the crystallinity of the monolayer surrounding the hole, as evidenced by the well-defined hexagonal diffraction patterns (Fig. 2C).

Subsequently, we mount a pore into a microfluidic flow cell, add a 1M saline solution (1M KCl, TE, pH 8.0) on both sides of the graphene membrane, and measure current-voltage (IV) curves from ion transport through the graphene nanopores (inset of Fig. 3). The resistance value (5.1 MΩ in the example of the inset of Fig. 3) and the linearity of the IV curve indicate that the current is consistent with ion flow through the pore and does not arise from electrochemical processes at the conductive graphene surface. Furthermore, samples with a graphene layer but without a nanopore exhibit a very high ionic resistance (>10 GΩ), which indicates that the graphene flake adheres well to the SiN surface and forms an insulating seal.

We measured IV curves for a number of pores ranging from 5 to 25 nm in diameter, both in graphene monolayers ($n = 6$) and multilayers ($n = 7$). Sample thickness is determined based on transmitted light intensity (2.3% reduction per layer). Fig. 3 shows the obtained resistances versus pore diameter, for both monolayers and multilayers with 2-8 layers (with a total thickness of 0.3 to 2.7 nm respectively). We find that the pore resistance scales with pore diameter as $R = \alpha/d^2$, with $\alpha = 2.4 \pm 0.2 \times 10^{-9}$ $\Omega m^2$.

Double-stranded DNA (dsDNA) can be driven electrophoretically through the nanopore and detected by monitoring the ion current. Upon addition of the 48.5 kb λ dsDNA (16 micron long) on one side of the pore and applying a voltage across the graphene membrane, a series of spikes is observed in the conductance traces (Fig. 4A). Each temporary drop in the measured conductance, $\Delta G$, arises from a single DNA molecule that translocates through the pore. As for conventional SiN nanopores[19], three characteristic signals are observed, corresponding to three types of translocation events: nonfolded (where the molecule translocates in a simple head-to-tail fashion), partially folded or fully folded molecules (where multiple pieces of the DNA are in the pore at the same time). Example events are shown in Fig. 4B. The events are color coded in black (nonfolded), red (partially folded) and blue (fully folded). From a large number ($n = 1222$) of such events, we obtain a histogram of conductance blockade levels $\Delta G$, as presented in Fig. 4C. Three peaks are visible, the first being the open-pore current at 0 nS (that is the baseline); the peak at ~1.5 nS which corresponds to one strand of DNA in the pore; and the peak at ~ 3 nS due to two strands of the same DNA molecule in the pore.

In addition to the event amplitude, we studied the translocation times of the events. A scatter plot of *ΔG* versus the time duration of the events is shown in Fig. 5, with the same color coding as used in Fig. 4B. Each dot in this diagram represents a single DNA translocation event. As expected, *ΔG* is twice as large for the folded events (2.9 ± 0.4 nS) compared to the nonfolded events (1.5 ± 0.4 nS). The blockade amplitude *ΔG* = 1.5 nS for nonfolded DNA in these graphene pores is quite similar in magnitude to that measured for pores of similar sizes in a 20 nm thick SiN membrane (1.4 ± 0.3 nS)[20]. The average translocation time is 2.7 ± 0.8 ms for the nonfolded DNA, a value that is notably larger than for solid-state nanopores in a 20 nm SiN membrane for which the translocation time is about 1.2 ms under the same applied voltage of 200 mV. Such a slower translocation is helpful for analytical applications such as local structure determination or sequencing.

The establishment of double-stranded DNA translocation to single-layer graphene nanopores represents an important step towards pushing the spatial-resolution limits of single-molecule nanopore analytics to subnanometer accuracy. This can be expected to impact single-nucleotide resolution in genomic DNA sequencing. Whereas this paper reports the first translocation of double-stranded DNA through graphene nanopores, future research will also explore single-strand DNA translocation and the engineering of graphene pores for sequencing.

**Acknowledgments.**

We thank M.-Y. Wu and Q. Xu for their assistance in TEM, H. Postma for discussions, V. Karas and K. van Schie for technical assistance, and the Netherlands Organisation for Scientific Research (NWO), the NanoSci-E+ Program, the Foundation for Fundamental Research on Matter (FOM), and the EC project READNA for funding.

**Methods.**

**Preparation of graphene samples for wedging transfer.**

We prepared graphene sheets on clean and freshly plasma-oxidized ($O_2$, Diener) Si/$SiO_2$ substrates by mechanical exfoliation of natural graphite (NGS Naturgraphit GmbH) with blue NITTO tape (SPV 224P). The plasma serves to make the substrate hydrophilic, which is needed for the wedging transfer. To render graphene monolayers visible, we used Si/$SiO_2$ wafers with a 90 nm thermally grown $SiO_2$ layer (IDB Technologies). We located the single and few layer graphene sheets under an optical microscope and identified the number of layers by their optical contrast as well as by Raman spectroscopy. Graphene flakes were transferred onto microfabricated Si/$SiO_2$/SiN chips described before[16]. We used cellulose acetate butyrate (Sigma-Aldrich) dissolved in ethyl acetate (30 mgs/mL) as the transfer polymer. Contrary to the design described by Krapf et al, prior the transfer of graphene, we etched the 20 nm thin SiN membrane using hot phosphoric acid (200 °C) for 45 minutes.

**Transmission electron microscopy and fabrication of nanopores in graphene.**

Nanopores were fabricated and imaged using a Cs-corrected Titan Cubed Supertwin/STEM FP5600/40 microscope operated at an accelerating voltage of 300 kV. An electron beam with a diameter of 15 nm at full width at half-maximum height and a beam density of $10^6$ electrons/(s·nm$^2$) was used for drilling. Gatan 2k x 2k CCD with binning 1 was used for image recording. Diffraction patterns were acquired with a beam size of 3 nm and a beam density of $10^5$ electrons/(s·nm$^2$). To remove contamination, samples were heated at 200 °C for at least 20 minutes prior to their insertion in the vacuum chamber of the microscope. After drilling, samples were stored in ethanol.

**Nanopore experiments.** For the electrical measurements, a membrane with a single graphene nanopore is mounted in a polyether ether ketone (PEEK) microfluidic flow cell and sealed to liquid compartments on either side of the sample. Measurements are performed in 1 M KCl salt solution containing 10 mM Tris-HCl and 1mM EDTA at pH 8.0 at room temperature. Ag/AgCl electrodes are used to detect ionic currents and to apply electric fields. Current traces are measured at 100 kHz bandwidth using a resistive feedback amplifier (Axopatch 200B, Axon Instruments), and digitized at 500 kHz. Before injecting dsDNA, the graphene-SiN-microchip was flushed with a 1 mg/mL solution of 16-mercaptohexadecanoic acid in 8:2 toluene/ethanol and additionally rinsed in respectively clean 8:2 toluene/ethanol and ethanol. This is expected to form a flat self-assembled monolayer on the graphene surface which demotes DNA adhesion[21]. dsDNA was unmethylated λ-DNA (20 ng/uL, Ref. No. D152A, lot no. 27420803, Promega, Madison,

USA). The event-fitting algorithm used to analyze and label the translocation events was the same as the one described before[20]. Only events exceeding 6 times the standard deviation of the open-pore root-mean-square noise are considered. Due to possible baseline fluctuations, we only considered events whose current before and after the event does not change more than 10% of the event amplitude. We additionally filtered the data at 10 kHz for better signal-to-noise ratio, and we discarded levels shorter then 200 μs.

**Figure 1.** Graphene Nanopores for DNA translocation (**A**) To-scale side-view illustration comparing DNA translocation through a SiN solid-state nanopore with that through a freestanding one-atom-thick graphene nanopore. (**B**) Optical micrographs depicting the transfer of graphene from Si/SiO$_2$ (left) onto a micro-fabricated silicon nitride chip containing a 5 μm hole (right). After the transfer by wedging, the flake entirely covers the hole (bottom image). (**C**) Raman spectrum of the flake on Si/SiO$_2$ before the transfer.

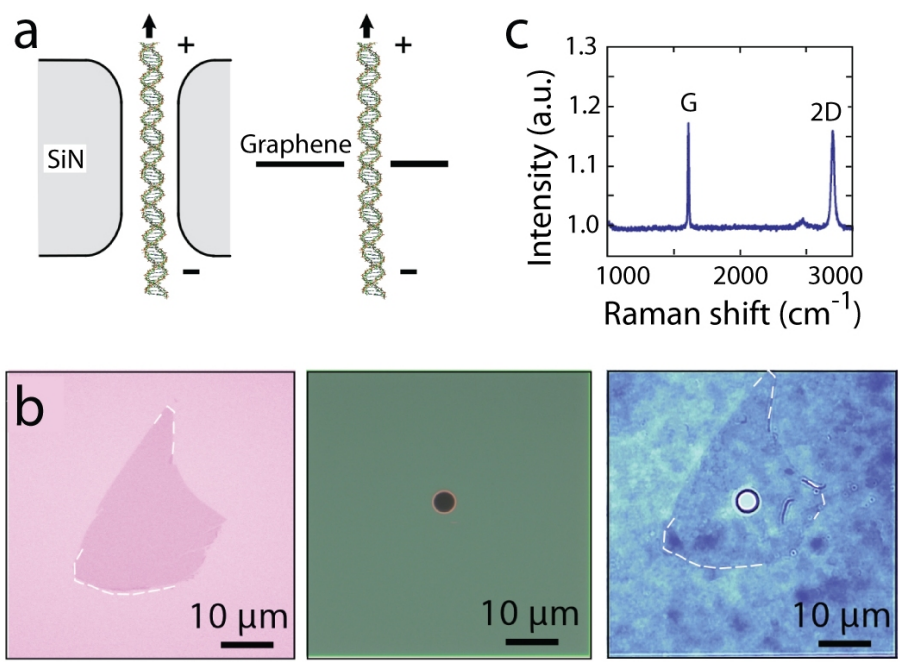

**Figure 2.** Drilling of graphene nanopores. (**A**) Transmission electron microscopy (TEM) of some nanopores drilled into multilayer graphene (**B**) TEM image (top) and diffraction patterns (bottom) across a 25 nm diameter pore in a monolayer of graphene. Numbers of the diffraction images indicate the spots where the patterns were recorded. (**C**) Diffraction patterns measured across the monolayer nanopore of panel B. The diffraction pattern was measured at three spots –indicated in panel B – with a 3 nm electron beam. The hexagonal lattice of diffraction spots is highlighted by the solid lines for clarity.

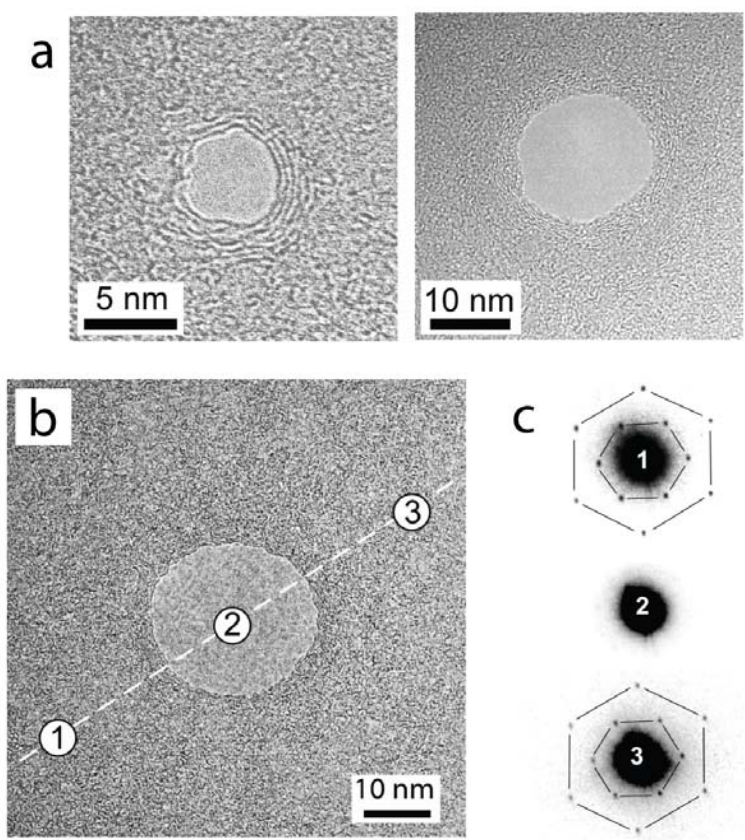

**Figure 3.** Nanopore resistances. Measured values of pore resistance versus diameter for a number of graphene nanopores (*n*=13). For each pore, the number of graphene layers is indicated by the number within the circle: 1 denotes graphene monolayers (blue); x denotes x layers of graphene (black). The solid line denotes a $1/d^2$ dependence. The inset shows an IV curve of a 22 nm nanopore in a graphene monolayer recorded in 1M KCl. A linear resistance of 5.1 MΩ is observed.

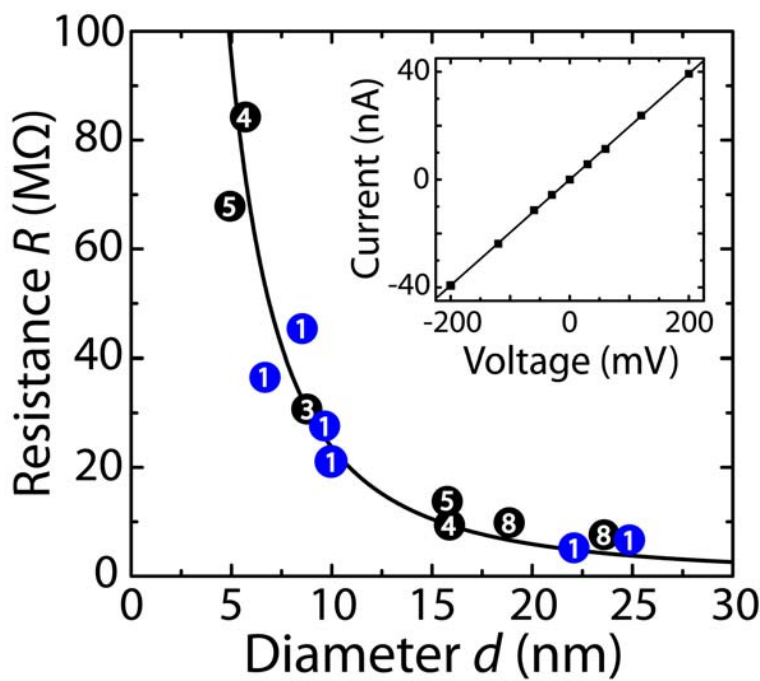

**Figure 4.** DNA translocation through a graphene monolayer. (**A**) Translocation of 48 kbp double-stranded λ-DNA across a 22 nm nanopore within a graphene monolayer, showing the baseline conductance (left) and blockade events upon addition of DNA (right). (**B**) Examples of translocation events of nonfolded (black), partially folded (red) and fully folded (blue) DNA molecules recorded at 200 mV. (**C**) Conductance histogram of 1222 translocation events, including 1 ms of open-pore conductance before and after the event. Note that counts in this histogram correspond to a single current measurement, not to a single event.

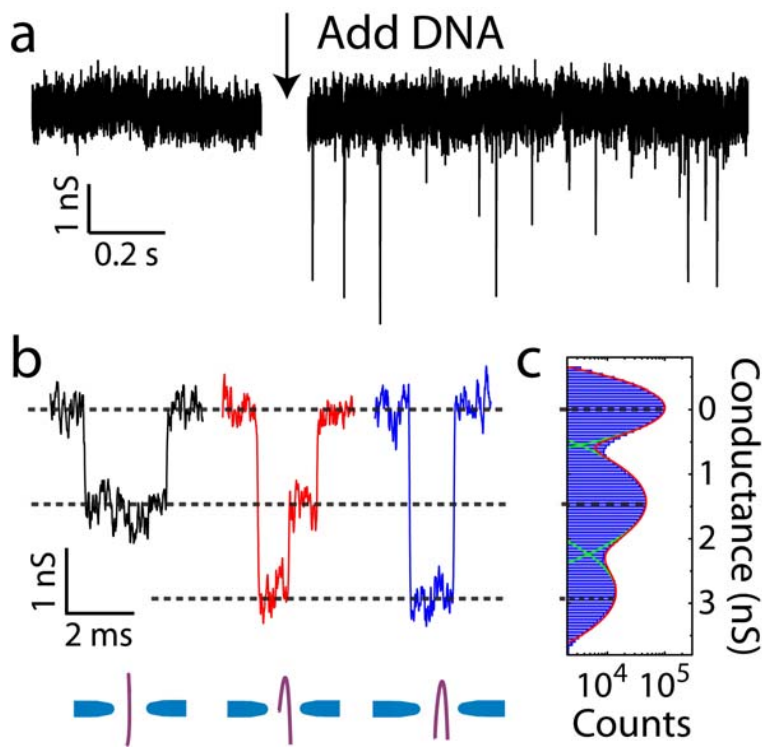

**Figure 5.** Event analysis. Scatter diagram of the amplitude of the conductance blockade versus translocation time with the accompanying histograms for the nonfolded and fully folded data at the top and the right. Color coding as in Fig. 4B. Each point in this scatter diagram corresponds to a single translocation event. For the labelling of events, each conductance data point within a translocation event (defined as an excursion of more than 6 times the standard deviation of the open-pore rms noise) is attributed to one of the peaks in the conductance histogram shown in Fig. 4C. The minimum required subsequent duration at one level is set to 100 μs, given by the rise-time resulting from the 10 kHz low-pass filtering.

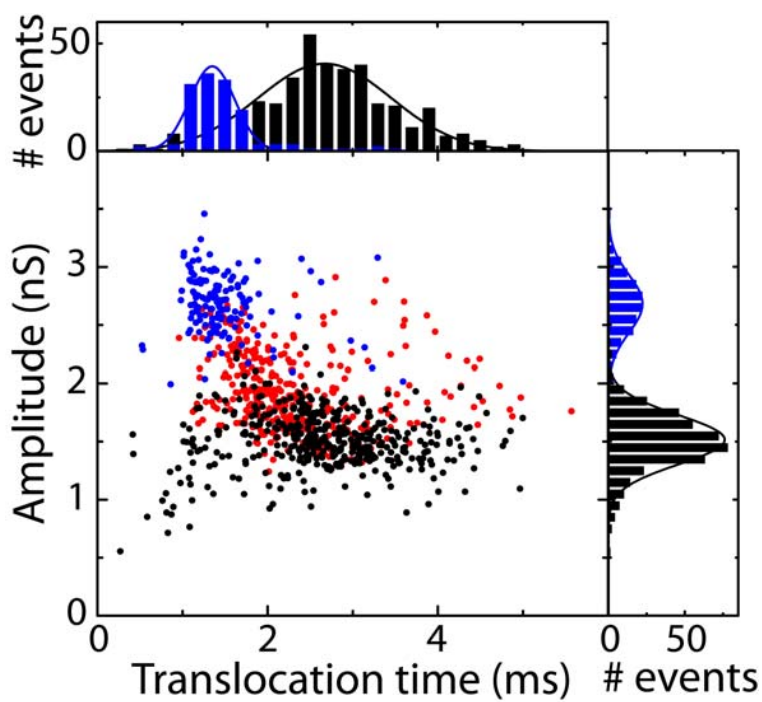